\documentclass[12pt]{article}
\usepackage{amstex,amssymb,cite}

\numberwithin{equation}{section}

\begin{document}

\begin{flushright}
SU-4240-675 \\
DSF-9/98 \\
IMSc 98/03/10 \\
SINP-TNP/98-01 \\
IC/98/37 \\
\end{flushright}

\begin{center}
\large{\bf Current Oscillations, Interacting Hall Discs and Boundary CFTs}

\bigskip
         A. P. Balachandran$^a$,\, 
         G. Bimonte$^b$\footnote{bimonte@@napoli.infn.it},\, 
         T. R. Govindarajan$^c$\footnote{trg@@imsc.ernet.in}, \\
	 K. S. Gupta$^d$\footnote{gupta@@tnp.saha.ernet.in}, \, 
         V. John$^e$\footnote{vjohn@@ictp.trieste.it} and \,
 	 S. Vaidya$^a$\footnote{sachin@@suhep.syr.edu} \\ 
$^a${\it Department of Physics, Syracuse University,}\\ 
    {\it Syracuse, N. Y. 13244-1130, U. S. A.}  \\ 
\smallskip
$^b${\it INFN, Sezione di Napoli, Napoli, Italy.} \\ 
\smallskip
$^c${\it Institute of Mathematical Sciences, Chennai, India.} \\ 
\smallskip
$^d${\it Theory Group, Saha Institute of Nuclear Physics, Calcutta, India.} \\ 
\smallskip
$^e${\it ICTP, Trieste, Italy.} 
\end{center}

\begin{abstract}
In this paper, we discuss the behavior of conformal field theories
interacting at a single point. The edge states of the quantum Hall effect
(QHE) system give rise to a particular representation of a chiral
Kac-Moody current algebra. We show that in the case of QHE systems
interacting at one point we obtain a ``twisted'' representation of the
current algebra. The condition for stationarity of currents is the same as
the classical Kirchoff's law applied to the currents at the interaction
point. We find that in the case of two discs touching at one point, since
the currents are chiral, they are not stationary and one obtains current
oscillations between the two discs. We determine the frequency of these
oscillations in terms of an effective parameter characterizing the
interaction. The chiral conformal field theories can be represented in
terms of bosonic Lagrangians with a boundary interaction. We discuss how
these one point interactions can be represented as boundary conditions on
fields, and how the requirement of chirality leads to restrictions on the
interactions described by these Lagrangians. By gauging these models we
find that the theory is naturally coupled to a Chern-Simons gauge theory in
$2+1$ dimensions, and this coupling is completely determined by the
requirement of anomaly cancellation.
\end{abstract}

\section{Introduction}
The motivation for studying interacting Hall systems is two-fold. Firstly
there are interesting experiments that can be performed on these systems
where a single point interaction comes about either from the experimental
geometry of the sample or due to an applied external voltage that induces
tunneling at one point between the edge currents of the quantum Hall system
\cite{cfksw}.  Secondly the effective field theories of these models
involve interesting theoretical concepts like topological field theory,
conformal field theory, quantization of anomalous gauge theories etc, and
studying these systems gives us a good testing ground for our understanding
of these quantum field theoretic ideas.  Related work on aspects of this
subject involves the Chern-Simons description of the quantum Hall effect
(QHE) \cite{sriwid}, this description is based on the fact that the low
lying excitations of the QHE system are chiral edge excitations that can be
described by a free field theory compactified at a radius that is related
to the Hall conductivity.  The fact that the effective field theory
corresponding to the bulk incompressible behavior of the QHE system is a
gauge theory requires one to understand various aspects of gauge theories
on finite geometries \cite{bbgs1,bbgs2}.  In particular the chiral scalar
field theory when gauged (coupled to electromagnetism) is anomalous and the
requirement of anomaly cancellation leads to a specific coupling of edge
and bulk theories.  Although much of this is by now standard, there are
still some matters that we need to clarify in order to generalize to the
case where we have ``touching discs'', that is two quantum Hall systems
with their respective discs, interacting through a single point on the
boundary.

In this paper we begin with a brief discussion of the origin of edge
currents in QHE systems in section 2. For simplicity, in this section, we
study QHE on a strip of finite width, though later we change the geometry
to that of a disc. (These different geometries differ in subtle details,
but the essential physics remains unaltered.)  In section 3, we discuss the
relationship of edge states to current algebras, using the fact that the
effective field theory of the low-lying excitations gives rise to a
fermionic theory and this in turn gives us a particular representation of
the Kac-Moody algebra on the Fock space of the fermionic theory. In section
4, we study how the current algebra is modified in the case of ``touching
discs'' by the interaction at the junction. We also discuss the edge
currents on the ``touching discs'', and show how one obtains current
oscillations between the discs. We then proceed to describe the
quantization of the edge theory.  And in section 5, we generalize the
situation to the case when there are several branches of edge
excitations. In section 6, we show that the edge theories can be described
in terms of a chiral bosonic Lagrangian with edge interactions. The edge
interactions lead to a free chiral bosonic theory with non-trivial boundary
conditions that we analyze in section 6. We also show in this section how
the theory is highly constrained and the interaction parameters are not
arbitrary. In section 7, we review the gauged chiral edge theory and its
coupling to a bulk Chern-Simons theory with a view to generalizing the
discussion to the case of touching discs.  In section 8 we show that the
interacting quantum Hall system can be described in terms of a twisted
complex boson that enables one to diagonalize the boundary interaction, and
that the requirement of anomaly cancellation determines the coupling of
this complex boson to a bulk Chern-Simons theory.

\section{Edge Currents on a Strip}
The importance of edge states for the quantum Hall effect (QHE) was first
recognized by \cite{halperin,taowu}. In this section, we will briefly review
the edge states for a system of non-interacting electrons moving in a
strong transverse magnetic field and their connection to current algebras.
So, consider an electron confined to move in the cylindrical strip
\begin{equation}
-X/2 \le x \le X/2, -Y/2 \le y \le Y/2
\end{equation}
with the lower and upper sides in the $y$ direction identified. The strip
is immersed in a strong magnetic field $B$, directed along the positive $z$
direction. The Hamiltonian for the electron then is
\begin{equation}
H= \frac {1}{2m} (\vec p + \frac {e}{c} \vec A)^2 + V(x)
\end{equation}
where $e$ is the absolute value of the charge of the electron, $m$ its
effective mass, $\vec A$ the vector potential for the
magnetic field and $V(x)$ the potential in the $x$- direction confining the
electrons to the strip. For simplicity, we assume for $V(x)$ the form of an
infinite-well potential localized at the edges. Moreover, it is convenient
to choose the gauge for $\vec A$ so that it is directed along the $y$-
direction and depends only on $x$:
\begin{equation}
A_x=A_z =0, \quad A_y=Bx. 
\end{equation}

The appropriate boundary conditions for the wave function are:
\begin{eqnarray}
\psi (x,-Y/2) &=& \psi(x,Y/2), \nonumber \\
\psi(\pm X/2,y) &=& 0.
\end{eqnarray}
The eigenfunctions then have the factorized form
\begin{equation}
\psi_k (x,y)=\frac {1}{\sqrt{Y}} e^{-iky}\psi_{k,\nu}(x), \quad k=\frac {2n
\pi} {Y}~,~n=0,\pm 1,\pm 2 \dots,
\end{equation}
where the meaning of $\nu$ will be discussed below.  The effective
Hamiltonian for $\psi_{k,\nu}(x)$ is
\begin{equation}
H_{eff}=-\frac {\hbar ^2}{2m} \frac{d^2}{dx^2} + \frac {m \omega_c^2}{2}
(x-l^2k)^2
\end{equation}
where $\omega_c$ is the cyclotron frequency and $l$ is the magnetic length:
\begin{equation}
\omega_c=\frac {eB}{mc}, \quad l^2=\frac {\hbar c}{eB}.
\end{equation}
In the conditions typical for the QHE,
\begin{equation}
B \approx 10T, \quad \omega_c \approx 10^{12} s^{-1},\quad l \approx 10^{-6}
cm.
\end{equation}

$H_{eff}$ is simply the Hamiltonian for a one- dimensional harmonic
oscillator centered around $X_k=l^2k$. Now, as long as $X/2-|X_k| \gg l$,
the boundary conditions at $x=\pm X/2$ can be neglected and the
eigenfunctions $\psi_{k,\nu}$ correspond to the usual harmonic oscillator
ones:
\begin{equation} 
\psi_{k,\nu}(x)= N_{\nu} e^{-(x-l^2k)^2/2 l^2} H_{\nu}(x/l - lk).
\end{equation}
Here $H_{\nu}$ is the Hermite polynomial of order $\nu$ and $\nu$ is also
its number of nodes between $-X/2$ and $X/2$. The corresponding energies
are
\begin{equation}
E_{k,\nu}=E_{\nu}= \hbar \omega_c (\nu+ \frac {1}{2}).  
\end{equation} 
We see that the energy depends only on $\nu$ and not on $k$ (Landau
levels).

That is not so when $X/2-|X_k| \approx l$ or when $|X_k| > X/2$: the
eigenfunctions $\psi_{k,\nu}$ can still be labeled by the number of nodes
they have between $-X/2$ and $X/2$, but their energies now depend also on
$k$. In particular, for a given $\nu$, the energies $E_{k,\nu}$ increase
with $|X_k|$ without a limit and it can be shown that
\begin{equation}
{\rm for} \quad |X_k|=\pm \frac {X}{2}, \quad E_{k,\nu}=\hbar \omega_c (2
\nu+ \frac {3}{2})
\end{equation}
while
\begin{equation}
{\rm for} \quad |X_k|-X/2 \gg l, \quad E_{k,\nu} \approx (l^2k- X/2)^2\frac
{e^2 B^2}{2 m c^2} .
\end{equation}

We will refer to the set of states having the same quantum number $\nu$ as
the $\nu$th Landau level.  Let us now consider the current, $I_{k,\nu}$,
carried in the $y$ direction by one of these states.  Use of the
Hellman-Feynman formula (see for example \cite{GalPas}) leads to the
following expression for $I_{k,\nu}$:
$$
I_{k,\nu}=\frac {1}{Y} \langle k,\nu |\frac {e}{m}(p_y + \frac{e}{c}A_y)
|k,\nu \rangle$$
\begin{equation}
=\frac {e}{\hbar Y} \langle k,\nu |\frac {\partial H}{\partial k}
|k,\nu \rangle=\frac {e}{Y \hbar} \frac {\partial E_{k,\nu}}{\partial k}.
\end{equation}

From here we see that only the states localized near the edges carry a non
vanishing current. These currents are diamagnetic. We will refer to them as
edge currents.

If the electrons are non-interacting, the multi-particle states are
obtained by placing each electron in a distinct Landau level (the spins are
aligned along the $z$- direction due to $B$). In the ground state, all the
levels up to the Fermi energy $E_F$, are filled. If $E_F$ lies between the
first and second Landau bands, a typical value for the total current
flowing in the proximity of the edge is
\begin{equation}
I_{edge} \approx 10^{-7}-10^{-8}~A.
\end{equation}

The question of the stability of these edge states in the presence of
impurities in the sample was discussed in the paper of Halperin we referred
to before. What happens is that, if the average distance between the
impurities is large compared to the magnetic length, the electrons
populating the edge states will simply go around them along equipotential
lines, and there will be no backscattering from the states on one edge to
the other. 
 
\section{Edge States and Current Algebras}
The edge states we have discussed in the previous section form a set of
states localized within a few magnetic lengths from the edge(s) of the
sample. At zero temperature, only the states below the Fermi energy will be
occupied.

Instead of a strip, let we will study a disc of circumference $L$, so that
we have only one edge, and suppose, for the sake of simplicity, that $E_F$
lies in the gap between the first and the second bulk Landau levels, so
that the occupied edge states belong only to the first Landau level.  If
one is interested only in the low energy excitations of the system, it is
then a good approximation to linearize the spectrum of the edge states
around the Fermi energy. The energy spectrum that we obtain is the spectrum
of a chiral, massless, complex Fermi field on the edge, propagating with a
velocity, $v_F$ equal to the group velocity at the Fermi energy. The
chirality of the electrons is determined by the direction of the external
magnetic field.  Let the boundary of the disc $\partial D$, of
circumference $L$, be coordinatized by its curvilinear coordinate $x$ (with
$x$ and $x+2\pi$ identified), increasing in the counterclockwise
direction. And assume that the direction of the external magnetic field is
such that the electrons near the edge move in the clockwise direction.
Such a theory is then described by the action
\begin{equation}
S=i \int dx dt(\psi_+^{\dagger} \dot {\psi}_+ - v_F
\psi_+^{\dagger}\partial_x \psi_+), 
\end{equation}
$$ 0 \le x \le L.$$
$\psi_+$ fulfills canonical anti-commutation relations:
\begin{equation}
[\psi_+^{\dagger}(x,t),\psi_+(x^{\prime},t)]_+= \hbar \delta(x-x^{\prime}).
\end{equation}
The equation of motion implied by this action is
\begin{equation}
\partial_- \psi_+=(\partial_t - v_F \partial_x)\psi_+=0
\end{equation}
which simply means that $\psi_+$ is a left- moving field:
\begin{equation}
\psi_+(x,t)=\psi_+(x+v_F t).
\end{equation}

Consider now the chiral current $J_+(x,t)$, associated with the field $\psi_+$,
\begin{equation}
J_+(x)=:\psi_+^{\dagger}(x) \psi_+(x):
\end{equation}
where the normal ordering is defined with respect to the annihilation and
creation operators for $\psi_+$. Due to one loop contributions, the
commutator of two such currents acquires an anomalous term (see for example
\cite{Ababro}), known as the Schwinger term:
\begin{equation}
[J_+(x,t),J_+(y,t)]=\frac {i \hbar ^2}{2 \pi} \delta ^{\prime}(x-y).
\label{curcom}
\end{equation}
The Fourier components 
\begin{equation}
K_n= \frac {1}{\hbar} \int dx J_+(x) e^{i2\pi n x/L}, \quad K_n ^{\dagger}
= K_{-n},\quad n\in {\mathbb Z}
\end{equation}
of the current fulfill the commutation relations defining an abelian
Kac-Moody (K-M) algebra,
\begin{equation}
[K_n,K_m]=n \delta_{n+m,0}.
\end{equation}
As is well known, the charge sectors of the theory are completely described
by the unitary, irreducible representations of this K-M algebra.  Moreover,
the Hamiltonian associated with the action can be expressed uniquely in
terms of the current $J_+(x)$ as
\begin{equation} 
H=\frac {2\pi v_F}{\hbar} \int dx
::J_+(x)J_+(x):: 
\end{equation}
 
The normal ordering $::\times \times \times::$ here is defined with respect
to the K-M generators:
\begin{eqnarray}
::K_n K_m:: &=& K_m K_n \quad {\rm if} \quad n > 0, \nonumber \\
::K_n K_m:: &=& K_n K_m \quad {\rm if} \quad n < 0.
\end{eqnarray}
\section{ Edge Currents on Two Touching Discs.}
In the previous section, we have discussed the edge currents flowing on the
rim of a disc in the presence of a strong transverse magnetic field.  A
question which arises naturally is how the behavior of these currents
changes if two discs are put in contact at one point. We will not attempt
here to analyze the detailed nature of the junction: the approach we will
pursue in this section is to see if the behavior of the edge currents can
be predicted on general grounds, starting from the current algebra
discussed in the previous section. What one expects is that the edge
currents will scatter at the junction, leading to a mixing of the edge
states for the two discs. Moreover, the scattering pattern should be such
that their chirality is preserved.

Consider two touching discs, $D_1$ and $D_2$, which we will assume, for
simplicity, to have the same circumference $L$.  Let the boundaries of
$D_i$, $i=1,2$ be coordinatized by $x_i: 0 \le x_i \le L$, measured
starting from the junction and increasing for both the discs in, say, the
anti-clockwise direction.

We now observe that, if a strong transverse magnetic field is turned on,
there will be, as before, states localized near the edges of the two discs
and that the corresponding edge currents will still fulfill a current
algebra of the form Eq.~(\ref{curcom}).  The problem, now, is that there is
a singular point, the junction, and we have to define the behavior of the
currents there. In order to do that, it is useful to observe that, fields
are operator valued distributions in quantum field theory, so that their
densities do not make sense as operators. Therefore, it is necessary to
smear the fields with suitable test functions in order to get well defined
self-adjoint operators. The test functions belong to a linear space and
encode the boundary conditions on the fields. In the case of two touching
discs, the test functions $\Lambda$, for our current algebra on the edge
are maps from the union of the boundaries of the two discs to real
numbers. As we shall soon see, they need not be continuous at the junction.

So, define the smeared currents
\begin{equation}
K(\Lambda)=\sum_i \int_{\partial D_i} dx_i \Lambda^i(x_i) J_+^i(x_i)
\end{equation}
where $\Lambda^i(x_i)$ and $J_+^i(x_i)$ denote the restrictions of
$\Lambda$ and $J_+$ to $\partial D_i$.  From Eq.(\ref{curcom}) we get, for
the commutator of two $K$'s,
\begin{equation}
[K(\Lambda),K(\bar {\Lambda})]=\frac {i \hbar^2}{2 \pi}\sum_k \int _{\partial
D_k} dx_k \Lambda^k d \bar {\Lambda}^k.
\label{commK}
\end{equation}
The boundary conditions on the smearing functions $\Lambda$ at the junction
must be such that the commutator Eq.(\ref{commK}) is antisymmetric with
respect to the interchange of $K(\bar {\Lambda})$ and $K({\Lambda})$. Thus
it must be that
$$0=[K(\Lambda),K(\bar {\Lambda})]+[K(\bar {\Lambda}),K(\Lambda)]=
\frac {i \hbar^2}{2 \pi}\sum_k \int _{\partial
D_k} dx_k (\Lambda^k d \bar {\Lambda}^k +\bar {\Lambda}^k d \Lambda^k)=$$

\begin{equation}
=\frac {i \hbar^2}{2 \pi}\sum_k (\Lambda^k(L) \bar {\Lambda}^k(L)-
\Lambda^k(0) \bar {\Lambda}^k(0)).
\end{equation}
This implies that for all test functions $\Lambda$,
\begin{equation}
\Lambda^{k}(L) = {\cal O}^k_l \Lambda^l (0)
\label{testfn}
\end{equation}
where ${\cal O}^k_l$ is some fixed orthogonal matrix characterizing the
junction.

Classically, the smearing functions can be identified with the possible
current distributions (at a given time). Thus the boundary conditions
Eq.~(\ref{testfn}) are nothing but the boundary conditions for the classical
currents at the junction:
\begin{equation}
J^k(L)={\cal O}^k_lJ^l(0).
\label{bcforJ}
\end{equation}

We see in this manner that a scattering matrix for the currents, at the
junction, arises naturally in this approach.  In the next section we will
study the physical implications of Eq.~(\ref{bcforJ}) on the behavior of the
currents circulating around the edges of two touching discs.  Here, we
notice only that the classical currents fulfilling the boundary conditions
Eq.~(\ref{bcforJ}) do not satisfy, in general, Kirchoff's law at the
junction. This implies, classically, that at different times there will
either be an accumulation or a depletion of electric charge at the
junction. Consequently, the total charge distributed on the rims of the two
discs $Q$, will not be conserved in time.  Actually this fact has, as we
will see, an even stronger consequence in quantum theory: we will find that
the corresponding charge operator, $\hat Q$ cannot be defined at all (in a
sense that will be explained later).
 
\subsection{Stationary Currents on the Edges of Two Touching Discs}
Before quantizing the current algebra Eq.~(\ref{commK}), in this section we
wish to present simple considerations on the behavior of the edge currents
on two touching discs.  The average current in a quantum edge-eigenstate
should correspond with a classical, stationary current distribution.  We
will study the stationary current distributions on the edges of two
touching discs, fulfilling the boundary conditions Eq.~(\ref{bcforJ}) .

In the stationary regime there cannot be accumulation of charge at the
vertex, and so besides the boundary condition Eq.~(\ref{bcforJ}), we must
have 
\begin{equation}
\sum_{i=1,2} J_+^i (L)=\sum_{i=1,2}J_+^i(0),
\label{kirchoff}
\end{equation}
which is nothing but the statement of Kirchoff's law at the junction.
The above equation, along with Eq.~(\ref{bcforJ}), implies that either
$J^i(L)=J^i(0)$ for $i=1,2$ or, $J^1(L)= J^2(0)$ and $J^2(L)=J^1(0)$.
 
\subsection{ Oscillating Edge Currents on Two Touching Discs}
As we discussed in the previous section the Kirchoff law condition
Eq.~(\ref{kirchoff}) is equivalent to demanding stationarity of currents.
However there are physical situations where we cannot require the currents
to be stationary in this sense.  In particular when we have a chiral edge
theory as in the quantum Hall effect we cannot have stationary edge
excitations, as the conditions for stationarity are incompatible with
chirality.  In fact, for the case of a single quantum Hall disc, there is
an edge current that propagates around the Hall disc with a frequency $\nu$
determined by the Fermi velocity $v_F$ where
\begin{equation}
\nu = \frac {v_F}{2 \pi L}.
\end{equation}

In this section we will see that in the absence of a stationarity
condition, we obtain transport of current from one disc to another and this
leads to current oscillations between the discs, with a frequency
characterized by the parameter $\alpha$ that is used to parameterize the
interaction between the currents on the two discs.

For two discs with equal circumference $L$, the frequency $\nu$ of the
oscillations is given by
\begin{equation}
\nu = \frac {\alpha v_F}{2 \pi L}
\label{frequency}
\end{equation}

These oscillations can be easily understood if one considers the
propagation of a classical impulse on the boundary. There are now two
cases, depending on the sign of det ${\cal O}$.  Let us consider now the
case when det~~${\cal O} = 1$ and hence,
\begin{equation}
{\cal O} = \left( \begin{array}{cc}
                       \cos \alpha & \sin \alpha \\
                       -\sin \alpha & \cos \alpha 
                  \end{array} \right).
\end{equation}

Suppose that, at time 0, there is a small wave packet on the left disc,
propagating in the clockwise direction.  When the impulse reaches the
junction, it splits according to the boundary conditions Eq.~(\ref{bcforJ})
into a transmitted impulse, propagating clockwise on the edge of the left
disc, and another transmitted impulse propagating in the clockwise
direction on the edge of the right disc.  There is no reflected impulse on
the left disc as the currents are chiral.  The transmitted impulse is equal
to $\cos \alpha$ times the initial one, and the scattered one to $-\sin
\alpha$ the initial one.  When these two impulses reach again the junction,
they again split according to Eq.~(\ref{bcforJ}) and then recombine and it
is easy to check that the resulting impulse on the left disc is equal to
$\cos 2\alpha$ times the initial pulse, while that on the right disc is
equal to $-\sin 2\alpha$ times the initial one.  After $n$ scatterings, the
impulse on the left disc is equal to $\cos n\alpha$ and that on the right
disc equal to $-\sin n\alpha$, times the starting impulse.  As discussed
above the oscillations arise basically because the relation
Eq.~(\ref{bcforJ}) also implies that after the $n$-th passage through the
junction
\begin{equation} 
J^i(L)= {\cal O}^{(n)} (\alpha) J^i(0). 
\end{equation} 

For simplicity we can take the case ${\cal O}^{(n)} (\alpha)=1$. This means
that the current oscillates between the two discs with a frequency given by
Eq.~(\ref{frequency}).  This also explains why there are no stationary
current distributions: they arise from taking time averages of these
oscillating currents, which of course vanish.  A similar analysis, for the
case det ${\cal O}=-1$, that is 
\begin{equation}
{\cal O} = \left( \begin{array}{cc}
                       -\cos \alpha & \sin \alpha \\
                       \sin \alpha & \cos \alpha 
                  \end{array} \right),
\end{equation}
gives a different result: the initial impulse on the first disc re-composes
after two scatterings since ${\cal O}^2 = {\bf 1}$. The average current on
each disc does not vanish and this explains the possibility of
non-vanishing stationary currents in this case.

\subsection{Quantization of the Edge Currents on Two Touching Discs}
In this section, we will quantize the edge currents on the boundaries of
two touching discs in the case det ${\cal O}=1$. An interesting aspect is
that the quantization cannot be done, in this case, directly in terms of a
Fermi field $\psi_+$ on the boundary, as was possible on a circle. The
reason is that there are no boundary conditions for the Fermi field which
imply the boundary conditions (\ref{bcforJ}) for the currents.  A simple
way to understand this is that the boundary conditions on $\psi$ which make
the Dirac Hamiltonian self-adjoint always imply Kirchoff's law at the
junction and we have seen before that our currents will not satisfy it in
general.

Thus, we have to quantize the currents directly. In order to do so, we
introduce a basis of smearing functions fulfilling the boundary conditions
(\ref{testfn}).  A convenient choice is
$$\Lambda_n \equiv \left(\begin{array}{c} 
			  \Lambda^1 \\ 
			  \Lambda^2
                          \end{array} \right) = 
	\frac {1}{\sqrt 2}\left(\begin{array}{c} 
					1 \\ 
					i \end{array} \right) 
e^{i(\alpha + 2\pi n)\frac{x}{L}},
$$

\begin{equation}
\Lambda^{*}_n \equiv \left(\begin{array}{c}
			    \Lambda^{1 *} \\
			    \Lambda^{2\ast}
		           \end{array} \right)=
\frac {1}{\sqrt 2} \left(\begin{array}{c}
			    1 \\ 
			   -i 
			 \end{array}\right)
e^{-i(\alpha + 2\pi n)\frac{x}{L}},
\end{equation}
$$n \in {\mathbb Z}, \quad 0 < \alpha < 2\pi.$$
We next introduce the corresponding smeared charges:
$$K_n \equiv \frac {1}{\hbar} K(\Lambda_n),$$
\begin{equation}
K_n^{\dagger} \equiv \frac {1}{\hbar} K(\Lambda_n^{\ast}).
\end{equation}
They fulfill the algebra
$$[K_n,K_m]=[K^{\dagger}_n,K^{\dagger}_m]=0.$$
\begin{equation}
[K_n,K^{\dagger}_m]= (\alpha + 2 \pi n) \delta _{n,m}.
\end{equation}
What we have got (modulo a normalization) is an infinite set of
annihilation and creation operators.  The vacuum, $|0>$, is defined by the
conditions
$$ K_n |0> =0, \quad n \geq 0,$$
\begin{equation}
K_n^{\dagger} |0> =0, \quad n < 0.
\end{equation}

As the Hamiltonian we assume the same expression, in terms of the currents
that we had on a disc:
\begin{equation}
H=\frac {2\pi v_F}{\hbar} \sum_k \int_{\partial D_k} dx_k
::J_+^k(x_k)J_+^k(x_k)::
\end{equation}
When expanded in terms of the generators $K_n$, this Hamiltonian coincides
with that of an infinite set of harmonic oscillators:
\begin{equation}
H= \frac {2 \pi v_F \hbar}{L} \sum_n :: K_n^{\dagger} K_n::
\end{equation}

Let us look now at the total charge $\hat Q$ of the discs:
\begin{equation}
\hat Q \equiv
\sum_k \int_{\partial D_k} dx_k J_+^k(x_k) = \hbar e^{i \pi/4}
(e^{-i \alpha}-1) \sum_n \frac {1}{\alpha + 2 \pi n} K_n + c.c.
\end{equation}
It can be checked that, when acting on the vacuum, $\hat Q$ creates a state
of infinite norm.
 
\section{ The Multichannel Case}
In this section we wish to examine the edge currents on two touching discs
when either the filling factor $n_1$ of the disc $D_1$ or that, $n_2$, of
disc $D_2$ or both are larger than 1. Because of the junction, there can
now be mixing between the several branches of edge currents circulating
at the boundaries of the two discs. We will consider, for simplicity,
the case of equal circumference $L$.

The generalization of the current algebra to this case is
\begin{equation}
\left \{K(\Lambda),K(\bar {\Lambda}) \right \}~=~ \sum_{I=1}^{n_1}
k^1_I\int_{\partial D_1}
\Lambda^1_I
d\bar {\Lambda}^1_I~+
\sum_{J=1}^{n_2} k^2_J\int_{\partial D_2}\Lambda ^2_Jd \bar {\Lambda}^2_J~,
\end{equation}
where $k^j_I$ are real numbers, reminiscent of the coefficients appearing in
Chern-Simons theory. The smearing functions $\Lambda$ and $\bar {\Lambda}$
can now be regarded as $N \equiv (n_1+n_2)$-dimensional column vectors:
\begin{equation}
\Lambda \equiv \left(\begin{array}{c}
		      \Lambda^1_1 \\ 
		      \vdots \\ 
		      \Lambda^1_{n_1} \\
		      \Lambda^2_1 \\ 
		      \vdots \\ 
		      \Lambda^2_{n_2}
		     \end{array}\right).
\end{equation}

If we now define the  $N \times N$ diagonal matrix $\cal K$,
\begin{equation}
{\cal K} \equiv diag(k^1_1,...,k^1_{n^1},k^2_1,...,k^2_{n^2}),
\end{equation}
we see that the antisymmetry of the Poisson bracket implies that the
smearing functions fulfill the boundary conditions
\begin{equation}
\Lambda (L)= A \Lambda (0)
\label{bcforLambda}
\end{equation}
where $A$ is some (fixed) $N \times N$
real matrix, which leaves $\cal K$ invariant:
\begin{equation}
A^{\dagger} {\cal K} A={\cal K}
\end{equation}
As in Eq.~(\ref{bcforJ}), we again identify the boundary conditions
Eq.~(\ref{bcforLambda}) with the boundary conditions for the edge currents
at the junction:
\begin{equation}
J(L)= AJ(0)
\end{equation}
where
\begin{equation}
I \equiv \left(\begin{array}{c}
		    I^1_1 \\ 
		    \vdots \\ 
		    I^1_{n_1} \\
                    I^2_1 \\ 
		    \vdots \\ 
		    I^2_{n_2}
	       \end{array}\right).
\end{equation}
The simplest case is when the matrix $A$ can be diagonalized. This happens,
for example, if we assume that $\cal K$ is proportional to the identity
matrix, because, then, $A$ is an orthogonal matrix. This is the case we
wish to consider now.  As is well known, if one regards an orthogonal
matrix $A$ as a linear transformation acting on a real $N$ dimensional
vector space $V$, it is always possible to decompose $V$ into invariant
two-dimensional and one-dimensional orthogonal subspaces,so that $A$ acts
as the identity on the one-dimensional ones and as a rotation on the
two-dimensional ones:
\begin{eqnarray}
V &=& \bigoplus _p V^{(2)}_p \bigoplus _q V^{(1)}_q, \quad dim \;
V^{(2)}_p=2, \quad dim \; V^{(1)}_q=1, \nonumber \\
A V^{(2)}_p &=& R(\alpha _p)V^{(2)}_p, \quad A V^{(1)}_q=V^{(1)}_q .
\end{eqnarray}
Notice that it follows from this that, if dim $A$ is odd, $q \ge 1$. The
behavior of the currents is then clear: for each subspace $V^{(1)}_q$,
there is a set of stationary currents ($I$ is along the direction of the
corresponding eigenvector of $A$) while for each $V^{(2)}_p$, there is a set
of oscillating currents, the frequency $\nu_p$ of the oscillations being
given by Eq. (\ref{frequency}) with $\alpha_p$ in the place of $\alpha$.
 
\section{Compact Scalar Field Theory with Boundaries}
The first issue we discuss in this section is regarding the proper
quantization of a compact chiral scalar field theory on manifolds with
boundaries.  Some of the issues that arise here are due to this latter
circumstance: we can now add boundary interactions to the theory,
interactions that will be required to preserve conformal invariance and
chirality.  The boundary equations of motion then lead to non-trivial
boundary conditions on the fields.  Thus apart from the standard Neumann
boundary conditions, we also have situations where there are Dirichlet
boundary conditions \cite{polchinski} or more general mixed boundary
interactions.

The Lagrangian for a single scalar field is given by
\begin{equation}
L= \frac{R^2}{2}\int d{\sigma}~ dt~~ \partial_{\mu} X \partial^{\mu} X + 
\int_{boundary} dt~ ({\rm boundary ~terms}).
\end{equation}

We have to deal with two cases here: the spatial geometry parameterized by
``$\sigma$'' can be a circle or a line segment. They correspond to closed
and open string theories respectively.  (Note: The closed string theory
corresponds to the situation where the Hall effect sample has a disc
geometry, while the open string theory is probably relevant if the geometry
is a rectangular bar/strip with length $l$.) If we are dealing with a
closed string, then the boundary term corresponds to a given but arbitrary
marked point on the circle.  The only constraints on the boundary terms are
that they should not spoil the conformal invariance of theory and should
also maintain the chirality.  The possible candidates for these terms for
the disc geometry are of the form
\begin{equation}
\int dt \left[A e^{ i\alpha X(0,t)} + Be^{-i \beta X(2\pi,t)}+ C [X(0,t)
 \partial_t X (2 \pi,t)- X(2\pi,t) \partial_t X(0,t)] \right]
\end{equation} 
where $\alpha$ and $\beta$ are fixed by requiring conformal invariance
i.e. the operators have conformal dimension one and they do not acquire
anomalous dimensions in perturbation theory.  The last term describes an
interaction between the fields at one point, physically this allows the
field $X$ to have discontinuities at this point. The exponential terms are
very interesting.  In a fermionized version they correspond to bilinears in
fermions and can describe tunneling of fermions.

In order to deal with compact chiral scalars we need to impose the
condition $X(\sigma + 2\pi, t)= X (\sigma , t) + 2\pi RN, N\in {\mathbb
Z}$, (which allows for winding modes), and the chirality condition $
\partial_z X =0 $, where $z= \sigma + it$.  The chirality constraint is
actually second class which means that it does not commute with itself for
different values of $\sigma$. But one can quantize the system using the
Dirac or some other prescription.

To make contact with the edge currents of the quantum Hall effect we will
have to gauge this theory. The required analysis is done in
\cite{chandar,bachsa1,bachsa2}, where it is shown that one has to deal with
an anomalous chiral theory living on the edge and the anomaly has to be
cancelled by an appropriate Chern-Simons term in the bulk.  The reason that
this chirality condition has to be dealt with delicately is because some
choices of boundary conditions will violate chirality and so there is a
tension between allowed boundary conditions and constraints.  Similarly the
gauge invariance of the boundary conditions has to be checked explicitly
too.

\subsection{Boundary Conditions}
Consider two scalar fields living on a circle of perimeter $L$, and
interacting only at the point $\theta = 0$ (which is identified with $L$).
We can write the action for this system as
\begin{equation}
S = \frac{R_1^2}{2}\int \partial_{\mu}X_1 \partial^{\mu}X_1 +
\frac{R_2^2}{2}\int \partial_{\mu}X_2 \partial^{\mu}X_2 +
B_{i,\alpha; j,\beta}\int X^i_\alpha \partial_t X^j_\beta dt,
\end{equation}
where $\alpha, \beta$ are 0 or $2\pi$ and $ X^i_\alpha= X^i(\alpha )$.  The
last term $B_{i,\alpha; j,\beta}$ represents a boundary interaction.  The
equations of motion are
\begin{eqnarray}
\Box X_1 &=& 0, \\
\Box X_2 &=& 0.
\end{eqnarray}
The boundary conditions can be compactly written as
\begin{equation}
R_i^2 \delta_{ij}(\sigma^3)_{\alpha \beta} \partial_\sigma X^j_\beta +
K_{i,\alpha; j, \beta}\partial_tX^j_\beta = 0, \quad i, \alpha \quad {\rm
fixed}
\end{equation}
where $K_{i,\alpha; j, \beta} = B_{i,\alpha; j,\beta} - B_{j,\beta;
i,\alpha}$ and $\sigma^3$ is the Pauli matrix.  If the fields $X_i$ are
chiral then we have
\begin{eqnarray}
\partial_\sigma X^1 + \epsilon_1 \partial_t X^1 &=& 0, \\
\partial_\sigma X^2 + \epsilon_2 \partial_t X^2 &=& 0, 
\end{eqnarray}
where $\epsilon_1, \epsilon_2 = \pm 1.$

Using the chirality conditions, we can eliminate $\partial_t X$ in favor of
$\partial_\sigma X$, and substitute in the boundary conditions. The
resulting equations can be compactly written as
\begin{equation}
\left[ \begin{array}{cccc}
        R_1 & -a \epsilon_1 R_1 & -b \epsilon_2 R_2 & -c \epsilon_2
R_2 \\
        -a \epsilon_1 R_1 & -R_1 & -d \epsilon_2 R_2 & -e \epsilon_2
R_2 \\
        b \epsilon_2 R_1 & d \epsilon_2 R_1 & R_2 & -f \epsilon_2
R_2  \\
        c \epsilon_1 R_1  & e \epsilon_1 R_1 & f \epsilon_2R_2 &
-R_2  
\end{array} \right] 
\left[ \begin{array}{c}
\partial_\sigma X_{2\pi}^1 \\ \partial_\sigma X_0^1
\\ \partial_\sigma X_{2\pi}^2 \\ \partial_\sigma X_0^2
\end{array} \right] \equiv MY = 0,
\label{chirality}
\end{equation}
where $K_{1,2\pi; 1,0}=aR_1^2, K_{1,2\pi; 2,2\pi}=bR_1R_2, K_{1,2\pi;
2,0}=cR_1R_2, K_{1,0; 2,2\pi}=dR_1R_2,
K_{1,0; 2,0}=eR_1R_2, K_{2,2\pi;2,0}=fR_2^2.$

Since $Y \neq 0$ in (\ref{chirality}), we require that det $M = 0$ in order
to obtain non-trivial solutions i.e.
\begin{equation}
[(af+cd-be)^2 + (b^2 + e^2 - c^2 - d^2)\epsilon_1 \epsilon_2 -
a^2 - f^2 + 1] = 0.
\label{detm}
\end{equation}

This seems to be a general case, but in order to study the oscillation
scenario, we make the simplifying assumption that $b = e = 0$ and also make
the assumption that $R_1= R_2 = R$.

The motivation for the choice $b = e = 0$ is that this is the simplest
boundary condition that gives rise to the kind of current oscillations
discussed in section 4.

Let us now study the case $\epsilon_1= \epsilon_2=1$, $b=e=0$ in greater
detail.  The boundary conditions in this case are

\begin{eqnarray}
\label{bc}
\partial_{\sigma} X^1_{2 \pi} &=&  a\partial_{\sigma} X^1 _{0} +
 c \partial_{\sigma} X^2 _{0}, \\
\partial_{\sigma} X^2_{2 \pi} &=&  -d \partial_{\sigma} X^1 _{0} +
f \partial_{\sigma} X^2 _{0}.
\end{eqnarray}
Comparing these equations with Eq.(\ref{bcforJ}) we find that $c=d=
sin \alpha$ and $a=f=cos \alpha$. It is trivial to check that
Eq.(\ref{detm}) is satisfied.

Now if we define the complex field $Z= X^1 + iX^2$ and its complex
conjugate $\bar Z =X^1 - iX^2$, then we find that Eq.(\ref{bc}) can be
rewritten as
\begin{eqnarray}
\label{bca}
\partial_{\sigma}Z(L) &=&  e^{i\alpha}\partial_{\sigma} Z(0), \\
\partial_{\sigma}\bar Z(L) &=&  e^{-i\alpha}\partial_{\sigma} \bar Z(0).
\end{eqnarray}
The solution for this boundary condition is that the field $Z$ (upto an
additive constant) is quasi-periodic and satisfies $Z(L)= e^{i\alpha}Z(0)$.
The mode expansions of $Z(\sigma)$ will involve quasi-periodic functions of
the form $e^{i(n+{\alpha\over 2\pi L}\sigma)}$.

\section{Chern-Simons Theory on a Disc}
In this section we will discuss some of the results that have already been
studied in the literature, \cite{emss,moosei,bcegs}, the main reason being
to set up the stage for generalizing to the case of interacting Hall
systems. In addition we think that there are still some interesting issues
to be understood here.  One such issue has to do with a proper mode
analysis. Careful mode analysis is indispensable already in classical
canonical formalism to resolve subtle mathematical problems; the latter
arise from the fact that there are fields supported both in the bulk and at
the edge.  The bulk and edge field theories are coupled and the nature of
the coupling is dictated by anomaly cancellation or gauge invariance.  One
thing that the mode analysis should allow one to see in detail is how the
Gauss law goes from being a second class constraint (as in the case of the
chiral Schwinger model) to a first class constraint.  This is another way
of saying that gauge invariance is restored by anomaly cancellation due to
the extra bulk contribution.

Let us rework our favorite example to illustrate our method of canonical
quantization. This is the usual Chern-Simons theory on a disc, along with a
boundary interaction term:
\begin{equation}
S = -\frac{k}{4\pi} \int_{D \times {\mathbb R}}{\cal A} d {\cal A}
 + \frac{k}{4\pi e} \int_{\partial D \times {\mathbb R}}d \phi{\cal A}.
\end{equation}

This is invariant under the gauge transformation
\begin{eqnarray}
        {\cal A} & \rightarrow & {\cal A} + d \lambda, \\
        \phi & \rightarrow & \phi - e \lambda.
\end{eqnarray}
The action does not contain the kinetic energy term for the $\phi$ field:
we will add it later. As it stands, this action describes a topological
field theory.

If we abstract the Lagrangian from the action, we notice that in terms of
the variables $z,\bar{z}$, $A_0(z,\bar{z})$ is a zero-form on the spatial
slice while the other components of ${\cal A}$ can be written in terms of a
one-form $A$. It is this one-form that we will expand in terms of an
appropriate basis set. The bulk Lagrangian is
\begin{equation}
        L_{bulk} =\frac{k}{4\pi}\int A\dot{A} - \frac{k}{2\pi}\int AdA_0.
\end{equation}
The wedge product is implicit in the above expression.

The edge Lagrangian is 
\begin{equation}
L_{edge} = \frac {k}{4\pi} \int_{\partial D} A_0 A_{\theta} + 
\frac{k}{4 \pi e}\int_{\partial D} [\dot{\phi}A_{\theta} -
\partial_{\theta}\phi A_0]. 
\end{equation}

The strategy for finding the modes for $A$ will be the same as the one
adopted in \cite{bcegs}. There are 4 kinds of 1-forms. The first two are
$h_n (z)$ and $\bar h_n (\bar z)$ where
\begin{equation}
h_n(z) = \frac{dz^n}{\sqrt{2\pi n}R^n}, \quad \bar{h}_n(\bar{z}) =
\frac{d\bar{z}^n}{\sqrt{2\pi n}R^n}, \quad n\geq 1.
\end{equation} 
Here $R$ is the radius of the disc. The other two modes are $\psi_{nM}$ and
$*\psi_{nM}$, where $*\psi_{nM}=N_{nM}dF_{nM}$. The $*$ is the Hodge
dual defined as in \cite{bcegs}. Also, $F_{nM}=e^{i n \sigma}G_{nM}$ where
$G_{nM}$ satisfies the Bessel equation
\begin{equation}
[\frac{d^2}{dr^2} + \frac{1}{r}\frac{d}{dr} + (\omega^2 -
\frac{n^2}{r^2})]G_{nM}(r) = 0,
\end{equation}
and the boundary condition $G_{nM}(R)= 0$, and $N_{nM}$ are chosen so that
$(*\psi_{nM}, *\psi_{n'M'}) = \delta_{n'n}\delta_{M' M}$. Thus
\begin{equation}
G_{nM} = J_{n}(w_{nM}r) \quad  n\in {\mathbb Z} \quad {\rm and}~ M \in
{\mathbb Z}_+ 
\end{equation}
and $\omega _{nM}$ are the solutions of $J_{n}(\omega_{nM}R)=0$.

The modes described above form a complete orthonormal set:
\begin{eqnarray}
(h_n, h_m) &=& (\bar{h}_n, \bar{h}_m) = \delta_{nm},  \nonumber \\
(\psi_{nM}, \psi_{n'M'}) &=& (*\psi_{nM}, *\psi_{n'M'}) =
\delta_{n'n}\delta_{M' M},
\end{eqnarray}
and
$(h_n, \bar{h}_m) = (h_n, \psi_{n'M'}) = (h_n, *\psi_{n'M'}) = 
(\psi_{nM}, *\psi_{n'M'}) = 0$ 
etc. The inner product $(\alpha, \beta)$ of one-forms $\alpha = \alpha_i
dx^i$ and $\beta = \beta_i dx^i$ is defined here as
\begin{equation}
(\alpha, \beta) =  \int \bar{\alpha}_i \beta_i d^2 x .
\end{equation}

Therefore we have, 
\begin{eqnarray}
A &=& \sum_{nM} [a_{nM}(t)\psi_{nM} + b_{nM}(t) *\psi_{nM}] +
                \sum_{n=1}^{\infty}\tilde\alpha_n(t)h_n + \nonumber\\ 
 &&\sum_{n=1}^{\infty}\tilde{\alpha}^{\dagger}_n(t)\bar{h}_n, \nonumber  \\ 
dA_0 &=& \sum_{nM}b_{0nM}(t) *\psi_{nM} +
                \sum_{n=1}^{\infty}\alpha_{0n}(t)h_n +
                \sum_{n=1}^{\infty}{\alpha}^{\dagger}_{0n}(t)\bar{h}_n 
\end{eqnarray}
where we have used the notation $a^{\dagger} _{nM}=a_{-n M}$, $b^{\dagger}
_{nM}=b_{-n M}$, and $b^{\dagger} _{0nM}= b_{0 -n M}$ to make the
expressions more concise.

The mode expansion for the field $\phi$ living on the boundary of 
the disc parameterized by the angular variable $0\leq\theta\leq 2\pi$ is 
\begin{equation}
\phi = \frac{1}{\sqrt {2\pi}}\sum_{n} \left( {\phi _{n}\over{\sqrt n}}e^{in
\theta} + {\phi^\dagger _{n}e^{-in \theta}\over{\sqrt n}} \right) + p
\theta + q, \quad n \in {\mathbb Z}_+ ,
\end{equation}
where $p, q$ correspond to the zero modes.  We will also need the mode
expansions for the fields at the boundary of the disc. They are
\begin{eqnarray}
A|_{\partial D} = a_0 d \theta &+& \sum_{n \geq 1} \left( \tilde{a}_n +
i \sqrt{\frac{n}{2 \pi}} \tilde{\alpha}_n \right) e^{i n \theta} d
\theta \nonumber \\
&+& \sum_{n \geq 1} \left( \tilde{a}_{-n} - i \sqrt{\frac{n}{ 2 \pi}}
\tilde{\alpha}^{\dagger}_{-n} \right) e^{-i n \theta} d\theta  \\ 
\nonumber
\end{eqnarray}
\begin{equation}
A_0|_{\partial D} = \alpha^{(0)}_0 + \sum_{n \geq 1} \left( \frac{\alpha
_{0n}}{\sqrt{2 \pi n}} e^{i n \theta} + \frac{\alpha^{\dagger}_{0n}}
{\sqrt{2 \pi n}} e^{-i n \theta} \right),
\end{equation}
where $\tilde{a}_n = \sum_{M}~ a_{nM}~ N_{nM}~~ \partial _{r}
J_{n}(\omega_{nM}r)|_{r=R}$, and $\alpha^{(0)}_0$ is a constant.

The Lagrangian  can be easily written in terms of these modes. 
\begin{eqnarray}
L = \sum_{nM}
&&{Ke \over 2} [-{a}_{nM}^\dagger  \dot{b}_{nM} + 
{b}_{nM}^\dagger \dot{a}_{nm} + i\left({\alpha_n}^\dagger \dot{\alpha_n}- 
\dot{\alpha_n}^\dagger {\alpha_n}\right)] \nonumber\\
&& +{Ke \over 2}
[2{a}^\dagger _{nM}b_{0nM} - 2i({\alpha_n}^\dagger {\alpha}_{0n}-
{\alpha_n} {\alpha}^\dagger_{0n})] \nonumber \\ 
&&-{Ke \over 2}[ \alpha^\dagger _{0n} ( a_n +i \alpha_n) +
\alpha_{0n} ( a^\dagger _n -i \alpha^\dagger _n)+\alpha_0 a_0]\nonumber \\
&&+{K\over 2}[ \dot{\phi_n} ( a_n^\dagger  -i \alpha^\dagger _n)
+ \dot{\phi_n}^\dagger  ( a_n  +i\alpha_n) + a_0 \dot q] \nonumber\\
&&-{K\over 2}[ i\alpha^\dagger _{0n}\phi_n -i\alpha _{0n}\phi^\dagger _n +
\alpha_{0}p]
\end{eqnarray}
where $K= \frac{k}{2\pi e}$ and $a_n = \sqrt{\frac{2\pi}{n}} \tilde{a}_n$, 
$\alpha _n = \sqrt{\frac{2\pi}{n}}\tilde{\alpha}_n$.

The conjugate momenta are:
\begin{eqnarray}
\Pi(a_{nM}) &=& {Ke\over 2} {b}^\dagger _{nM}, \nonumber \\
\Pi(b_{nM}) &=& -{Ke\over 2} {a}^\dagger _{nM},\nonumber \\
\Pi(\alpha_{0n}) &=& 0 = \Pi(\alpha^\dagger _{0n}), \nonumber\\
\Pi(b_{0nM})&=& 0 \nonumber\\
\Pi(\alpha_n) &=& {Kei\over 2}{\alpha_n}^\dagger ,\nonumber\\
\Pi(\dagger{\alpha_n}) &=& {-Kei\over 2}{\alpha_n},\nonumber\\
\Pi({\phi_n}) &=& K/2( a_n^\dagger  -i \alpha^\dagger _n) \nonumber\\
\Pi({\phi_n}^\dagger) &=& K/2( a_n  +i\alpha_n).
\end{eqnarray}

All the above are primary constraints. Doing the constraint analysis we
obtain further secondary constraints. They are
\begin{eqnarray}
a_{nM} &\sim & 0 \nonumber\\
e\alpha_{n} -\phi_n &\sim& 0 \nonumber\\
e\alpha^{\dagger}_{n}-\phi^{\dagger}_n &\sim& 0. 
\end{eqnarray}

The Poisson bracket matrix of the constraints has a zero determinant
indicating the first class nature of some constraints.  The set of
constraints is
\begin{eqnarray}
\chi_1 \equiv \Pi(\alpha_n) -{iKe\over 2}\alpha_n^\dagger = 0,
& & \chi_2 \equiv \Pi(\alpha_n^\dagger) +{iKe\over 2}\alpha_n =0, \nonumber\\
\chi_3 \equiv e\alpha_n - \phi_n=0, & & \chi_4 \equiv e \alpha_n^\dagger - 
\phi_n^\dagger =0, \nonumber\\
\chi_5 \equiv e \Pi(\phi_n) + {iK\over 2}\phi_n^\dagger = 0,
& & \chi_6 \equiv e\Pi(\phi_n^\dagger) - {iK\over 2}\phi_n = 0.
\end{eqnarray}

In writing this set of equations we have used $e\alpha_n - \phi_n=0$ in the
constraint for the momenta of $\phi$ which amounts to taking some linear
combinations of constraints and calling that as a new constraint.

We use the Poisson brackets 
\begin{equation}
\{\alpha_n,\Pi(\alpha_n)\} = \{\phi_n, \Pi(\phi_n)\} =1.
\end{equation}

The Poisson bracket matrix is 
\begin{equation} 
[\{ \chi_i, \chi_j \}] = 
\left(
\begin{array}{cccccc}
 0&-iKe&-e&0&0&0\\ 
 iKe&0&0&-e&0&0\\
 e&0&0&0&-e&0\\ 
 0&e&0&0&0&-e \\ 
 0&0&e&0&0&iKe \\
 0&0&0&e&-iKe&0 
\end{array}\right)
\label{PBmatrix}
\end{equation}

This matrix has zero determinant.  The eigenvectors for zero eigenvalue
are (in the same basis of constraints that the matrix is written in
(\ref{PBmatrix}))
\begin{equation}
\begin{array}{cccccc}
(0, 1, -iK, 0, 0, 1), \\ 
(1, 0, 0, iK, 1, 0).
\end{array}
\label{eigen}
\end{equation}
These zero eigenvectors correspond to linear combinations of first class
constraints that are the generators of gauge transformations on the edge of
the disc.  Doing the constraint analysis, we find that the Lagrange
multipliers corresponding to the second class constraints get fixed, and
the Hamiltonian can be written as a sum of constraints. Since this theory
is completely topological, the Hamiltonian is weakly zero as expected. 

The second class constraints force us to construct Dirac brackets for
quantizing the theory. The Dirac brackets for the gauge invariant edge
observables are
\begin{equation}
\{ \alpha_n + \frac{1}{e} \phi_n, 
{\alpha}^\dagger _m +\frac{1}{e} \phi_m^{\dagger} \}^{DB} = 
\frac{-4 \pi i}{k}\delta_{nm}.
\end{equation}

\section{Dynamics for the Edge Field}
We can now include the kinetic energy term for the scalar field at the
edge. This is just an extra term in the Lagrangian of the form
\begin{equation}
\frac{k}{8 \pi e^2} \int D_\mu \phi D^\mu \phi.
\end{equation}
The analysis now, though identical in spirit to the previous one, differs
in detail.

The new Lagrangian $L'$ can be easily written in terms of these modes,
\begin{eqnarray}
L' = \sum_{nM}
&&{Ke \over 2} [-{a}_{nM}^\dagger  \dot{b}_{nM} + 
{b}_{nM}^\dagger \dot{a}_{nm} + i\left({\alpha_n}^\dagger \dot{\alpha_n}- 
\dot{\alpha_n}^\dagger {\alpha_n}\right)]
\nonumber\\
&& +{Ke \over 2}
[2{a}^\dagger _{nM}b_{0nM} - 2i({\alpha_n}^\dagger {\alpha}_{0n}-
{\alpha_n} {\alpha}^\dagger_{0n})] \nonumber \\ 
&&-{Ke \over 2}[ \alpha^\dagger _{0n} ( a_n +i \alpha_n) +
\alpha_{0n} ( a^\dagger _n -i \alpha^\dagger _n)+\alpha_0 a_0]\nonumber \\
&&+{K\over 2}[ \dot{\phi_n} ( a_n^\dagger  -i \alpha^\dagger _n)
+ \dot{\phi_n}^\dagger  ( a_n  +i\alpha_n) + a_0 \dot q] \nonumber\\
&&-{K\over 2}[ i\alpha^\dagger _{0n}\phi_n -i\alpha _{0n}\phi^\dagger _n +
\alpha_{0}p]\nonumber\\
&& +{K\over 2 en}(\dot{\phi_n}^\dagger + e \alpha^\dagger _{0n})
(\dot{\phi_n} + e \alpha_{0n}) \nonumber\\
&&+{Kn\over 2 e}(e a^{\dagger} _{n} -i \phi^{\dagger} _n -ie 
\alpha^{\dagger} _n)(e a_{n} -i \phi_n -ie \alpha_n)
\end{eqnarray}
where $K= \frac{k}{2\pi e}$ and $a_n = \sqrt{\frac{2\pi}{n}} \tilde{a}_n$, 
$\alpha_n = \sqrt{\frac{2\pi}{n}}\tilde{\alpha}_n$.

The following conjugate momenta $\Pi(\chi)$ of $\chi \; (= a_{nM}\;
\text{etc})$ are all constrained:
\begin{eqnarray}
\Pi(a_{nM}) &=& {Ke\over 2} {b}^\dagger _{nM}, \nonumber \\
\Pi(b_{nM}) &=& -{Ke\over 2} {a}^\dagger _{nM},\nonumber \\
\Pi(\alpha_{0n}) &=& 0 = \Pi(\alpha^\dagger _{0n}), \nonumber\\
\Pi(b_{0nM})&=& 0 \nonumber \\
\Pi(\alpha_n) &=& {Kei\over 2}{\alpha_n}^\dagger ,\nonumber\\
\Pi({\alpha_n}^{\dagger}) &=& {-Kei\over 2}{\alpha_n}.
\end{eqnarray}

We also have the canonical momenta for the field $\phi$:
\begin{eqnarray}
\Pi({\phi_n}) &=& K/2( a_n^\dagger  -i \alpha^\dagger _n) +
{K\over 2 en}(\dot{\phi_n}^\dagger + e \alpha^\dagger _{0n}), \nonumber \\
\Pi({\phi_n}^\dagger) &=& K/2( a_n  +i\alpha^\dagger _n)+
{K\over 2 en}(\dot{\phi_n} + e \alpha_{0n}). 
\end{eqnarray}

The Hamiltonian can be found:
\begin{eqnarray}
H &=& {2en\over k} \Pi_n ( \phi ^\dagger _n) \Pi_n ( \phi  _n) 
+ ien[\Pi_n ( \phi ^\dagger _n) \alpha ^\dagger _n  -\Pi_n ( \phi _n) 
\alpha _n] \nonumber \\
&& +{Ken\over2} \alpha ^\dagger _n \alpha _n 
+ \alpha^\dagger _{0n} (-e \Pi ^\dagger _n + {iK\over 2}\phi _n)
+\alpha _{0n} (-e \Pi _n - {iK\over 2}\phi^\dagger  _n) \nonumber \\
&& + ({\rm constraints} )
\end{eqnarray}

There are two first class constraints that generate gauge transformations
namely $\Pi (\alpha^\dagger _{0n}) ,\Pi (\alpha _{0n}) $ .
There are no first class generators corresponding to the zero 
eigenvectors that we found in Eq~(\ref{eigen}) when there was no
kinetic energy term for the chiral boson at the edge.
There are four second class constraints $ \Pi (\alpha^\dagger _{n}), \Pi
(\alpha _{n}) - iKe \alpha^ \dagger _n , e \Pi _n +{iK \over 2} \phi
^\dagger _n , e \Pi ^\dagger _n -{iK \over 2} \phi _n $. These can be
eliminated using Dirac brackets and imposing the second class constraints
strongly.

The non-vanishing Dirac brackets are:
\begin{eqnarray}
\{ \alpha_n, {\alpha}^\dagger _m \}^{DB} &=& \frac{-2 \pi i}{k}\delta_{nm},
\nonumber \\
\{ \phi_n, \Pi(\phi_m) \}^{DB} &=& \frac{1}{2} \delta_{nm}, \nonumber \\
\{ \phi_n, (\phi^\dagger _m) \}^{DB} &=& \frac{2\pi i e^2}{k} \delta_{nm},
\nonumber \\
\{ \Pi(\phi_n), \Pi^\dagger (\phi_m) \}^{DB} &=& \frac{-ik}{8\pi e^2} 
\delta_{nm}.
\end{eqnarray}

We can now write the Hamiltonian as
\begin{eqnarray}
H &=& {nK\over 4e} [ (p_n +ie \alpha ^\dagger _n)^\dagger 
(p_n +ie \alpha ^\dagger _n) \nonumber \\
&+& (\phi _n +e\alpha _n)^\dagger (\phi_n +e\alpha _n)] + h.c.  
\end{eqnarray} 
where ${2e\Pi _n \over K}= p_n$

In this section, we have just discussed the quantization of a Chern-Simons
theory on the disc coupled to a scalar field on the edge.  It can be
extended to the case of the touching discs where we now will have two
Chern-Simons fields. Due to the edge interaction we find it convenient to
work in terms of a complex combination of these real fields. We then find
(as discussed in section six) that due to the interaction at the contact
point the modes are twisted.  The quantization of the twisted chiral
complex scalar field coupled to a complex Chern-Simons field can be done
using an analysis similar to the one in this section. We will summarize the
results in the next section.

\section{Touching Discs}

Instead of two real fields $X^1(\sigma, t)$ and $X^2(\sigma, t)$ on the
boundary, we can work with a single complex field $Z=X^1 + iX^2$. In terms
of this $Z$, the boundary conditions can be written as
\begin{eqnarray}
\partial_{\sigma}Z_{2\pi} &=& e^{i2\pi \alpha}\partial_{\sigma} Z_0, \\
Z_{2\pi} &=& e^{i2\pi \alpha}Z_0
\end{eqnarray}
and complex conjugates of the above equations. Thus $Z$ is
``quasi-periodic''.

With $A=A^1 +i A^2$, we write the action as 
\begin{equation}
        S = -k \int_{D \times {\mathbb R}}(\bar{A}dA + Ad\bar{A})
        +\frac{k}{e}\int_{\partial D \times {\mathbb R}}(d\bar{Z}A +
        dZ\bar{A}),
\label{action}
\end{equation}
where $\bar{A}$ is the complex conjugate of $A$.

The bulk Lagrangian is  
\begin{equation}
        L_{bulk} = k\int(\bar{A}\dot{A} + A\dot{\bar{A}}) -
        k\int(\bar{A}dA_0 + Ad\bar{A_0}) + k
        \int(d\bar{A_0}A + dA_0\bar{A}).
\label{bulkL}
\end{equation}

Similarly, the edge Lagrangian is 
\begin{equation}
        L_{edge} = -k\int_{\partial D}(\bar{A_0}A + A_0\bar{A}) +
        \frac{k}{e}\int[(\dot{\bar{Z}}A_{\sigma} + \dot{Z}\bar{A_{\sigma}})
        - (\partial_{\sigma}\bar{Z} A_0 + \partial_{\sigma}Z \bar{A_0})].
\label{edgeL}
\end{equation}

We need a convenient mode expansion for $A$, $dA_0$, $Z$ and their complex
conjugates. Also, $A$, $dA_0$ and $Z$ must be of the same quasi-periodicity
for the action to be well-defined.  

Here, any 1-form in the bulk can be written as a linear combination of the
following 4 types of one-forms: 
\begin{eqnarray}
h_n(z) &=& \frac{dz^{n+\alpha}}{\sqrt{2\pi(n+\alpha)}R^{n+\alpha}},
        n\geq 0, \\ 
\bar{g}(z) &=&
        d\frac{\bar{z}^{n-\alpha}}{\sqrt{2\pi(n-\alpha)}R^{n-\alpha}},
        n\geq 1.   
\end{eqnarray}
The other two modes are $\psi_{nM}$ and $*\psi_{nM}$, where
$*\psi_{nM}=N_{nM}dF_{nM}$. Here $F_{nM}=e^{i(n+\alpha)\sigma}G_{nM}$, and
$G_{nM}$ satisfies the Bessel equation
\begin{equation}
[\frac{d^2}{dr^2} + \frac{1}{r}\frac{d}{dr} + (\omega^2 -
\frac{(n+\alpha)^2}{r^2})]G_{nM}(r) = 0,
\end{equation}
while $N_{nM}$ are so chosen that $(*\psi_{nM}, *\psi_{nM}) = 1$. Thus
\begin{eqnarray}
G_{nM} &=& J_{n+\alpha}(w_{n+\alpha,M}r), n\geq 0, \\
       &=& J_{-(n+\alpha)}(w_{-(n+\alpha),M}r), n\leq -1
\end{eqnarray}
and $G_{nM}(R)=0$.

Therefore we have, 
\begin{eqnarray}
        A &=& \sum_{nM}[a_{nM}(t)\psi_{nM} + a^{(*)}_{nM}(t)*\psi_{nM}] \\
          & &  + \sum_{n=0}^{\infty}\alpha_n(t)h_n +
                 \sum_{n=1}^{\infty}\bar{\beta}_n(t)\bar{g}_n,  \\
        dA_0 &=& \sum_{nM}a^{(*)}_{0nM}(t) *\psi_{nM} +
                \sum_{n=0}^{\infty}\alpha_{0n}(t)h_n +
                \sum_{n=1}^{\infty}\bar{\beta}_{0n}(t)\bar{g}_n 
\end{eqnarray} 
and
\begin{equation}
        Z=\sum_{n=-\infty}^{+\infty}\frac{z_n}{\sqrt{2\pi |n + \alpha|}}
                e^{i(n+\alpha)\sigma}.       
\end{equation}
We also need $A|_{\partial D}$ and $A_0|_{\partial D}$. We find them to be
\begin{eqnarray}
        A|_{\partial D} &=& \sum_{nM}a_{nM}f_{nM} e^{i(n+\alpha)\sigma}
                d \sigma + \sum_{n=0}^{\infty} i\alpha_n
                \sqrt{\frac{n+\alpha}{2\pi}} e^{i(n+\alpha)\sigma}d
                \sigma \\
             &&+\sum_{n=1}^{\infty} i\bar{\beta}_n \sqrt{\frac{n-\alpha}{2\pi}}
                e^{-i(n-\alpha)\sigma}d \sigma  \\
        A_0|_{\partial D} &=& \sum_{n=0}^{\infty} i\alpha_{0n}
                \sqrt{\frac{n+\alpha}{2\pi}} e^{i(n+\alpha)\sigma}d \\
             &&\sigma +
                \sum_{n=1}^{\infty} i\bar{\beta}_{0n}
                \sqrt{\frac{n-\alpha}{2\pi}} e^{-i(n-\alpha)\sigma}d \sigma 
\end{eqnarray}
where 
\begin{eqnarray}
        f_{nM} &=& N_{nM}w_{n+\alpha,M}R J'_{n+\alpha}(w_{n+\alpha,M}R),
                n \geq 0,  \\
        &=& N_{nM}w_{-(n+\alpha),M}R J'_{-(n+\alpha)}(w_{-(n+\alpha),M}R),
                n\leq -1.
\end{eqnarray}
For future convenience, we define 
\begin{equation}
        a_n \equiv \sum_{M}a_{nM}f_{nM}
\end{equation}

The full Lagrangian $L = L_{bulk}+L_{edge}$ can be written in terms of the
modes as

\begin{eqnarray}
L &=& k\sum_{nM}(a^{(*)}_{nM}
        \dot{\bar{a}}_{nM}-a_{nM}\dot{\bar{a}}^{(*)}_{nM}) 
        + k\sum(i \bar{\alpha}_n \dot{\alpha}_n + i \bar{\beta}_n
        \dot{\beta}_n) \\
   && + 2k\sum_{nM}a_{nM}\bar{a}^{*}_{0nM} + k\sum(i \alpha_n
        \bar{\alpha}_{0n} + i\beta_n \bar{\beta}_{0n}) \nonumber  \\ 
        & & - k\sum(\frac{\bar{\alpha}_{0n}a_n}{\sqrt{2\pi(n + \alpha)}} +
        \frac{\beta_{0n}a_{-n}}{\sqrt{2\pi(n-\alpha)}}) + \frac{k}{e}
        \sum_n \frac{\dot{z}_n a_n}{\sqrt{2\pi |n + \alpha|}}\nonumber\\ 
    && - \frac{k}{e} \sum i(\dot{z}_n \bar{\alpha}_n - 
        \dot{z}_{-n} \beta_n) \nonumber 
         - \frac{k}{e} \sum i(z_n\bar{\alpha}_{0n} -
        z_{-n}\beta_{0n}) + \rm{c.c.}
\end{eqnarray}

We can now write the momenta conjugate to the various fields $q_n$, using
$\Pi(q_n) \equiv \displaystyle{\frac{\partial L}{\partial \dot{q}_n}}$:
\begin{eqnarray}
        \Pi(a_{nM}) &=& k \bar{a}_{nM}^{(*)}, \\
        \Pi(a^{(*)}_{nM}) &=& -k \bar{a}_{nM}, \\
        \Pi(\alpha_n) &=& i k \bar{\alpha}_n, \\
        \Pi(\beta_n) &=& i k \bar{\beta}_n, \\
        \Pi(a^{(*)}_{0nM}) &=& 0, \\
        \Pi(\alpha_{0n}) &=& 0,\\
        \Pi(\beta_{0n}) &=& 0, \\
        \Pi(z_n) &=& \frac{k}{e} \frac{\bar{a}_n}{\sqrt{2 \pi |n+\alpha|}} -
                \frac{ik}{e}\bar{\alpha}_n, n \geq 0, \\
                &=& \frac{k}{e} \frac{\bar{a}_n}{\sqrt{2 \pi |n+\alpha|}} -
                \frac{ik}{e} \beta_{-n}, n \leq -1, \\ 
        \Pi(\bar{q}_{n}) &=& \bar{\Pi}(q_n).   
\end{eqnarray}

The Hamiltonian $H$ can be calculated from the Lagrangian to be 
\begin{eqnarray}
        H &=& -2k\sum_{nM} \bar{a}_{nM} a^{(*)}_{0nM} + k\sum
        (\frac{a_n}{2\pi(n+\alpha)} -i\alpha_n
        + \frac{i}{e} z_n)\bar{\alpha}_{0n} + \nonumber \\
        & & k\sum(\frac{a_{-n}}{2\pi(n-\alpha)} + i\bar{\beta}_{n} +
        \frac{i}{e}z_{-n})\beta_{0n} + \rm{c.c.}
\end{eqnarray}

The Poisson brackets follow their form $\{ q_n, \Pi(q_m) \} = \delta_{nm}$
for the various fields. These relations in turn imply that

\begin{eqnarray}
        \{ a_{nM}, \bar{a}^{(*)}_{n'M'} \}
                &=& \frac{1}{k} \delta_{nn'} \delta_{MM'}, \\  
        \{ \alpha_n, \bar{\alpha}_m \} &=& \frac{-i}{k}\delta_{nm}
\end{eqnarray}
and so on.

Since $\Pi(a^{(*)}_{0nM})=0$, we insist that $\dot{\Pi}(a^{(*)}_{0nM}) =
\{\Pi(a^{(*)}_{0nM}), H\}$ annihilates any physical state $|\phi>$. This,
and other such relations imply

\begin{eqnarray}
        a_{nM}|\phi> &=& 0, \\
        \bar{a}_{nM}|\phi> &=& 0, \\
        (\bar{\alpha}_n - \frac{1}{e}\bar{z}_n)|\phi> &=&
                0, n\geq 0\\
        (\alpha_n - \frac{1}{e}z_n)|\phi> &=& 0, n \geq 0 \\
        (\bar{\beta}_{-n} + \frac{1}{e}z_n)|\phi> &=&
                0, n \leq -1 \\ 
        (\beta_{-n} + \frac{1}{e}\bar{z}_n)|\phi> &=&
                0, n \leq -1
\end{eqnarray}

The last four constraints do not commute amongst each other. So quantizing
this system of constraints requires us to define the Dirac bracket. Again
we can calculate the Dirac bracket of various operators with each
other. The interesting ones are
\begin{eqnarray}
\{ q_n, \bar{q}_m \} &=& \frac{-4i}{k}\delta_{nm}, \\
\{ z_m, \bar{z}_n \} &=& \frac{-4ie^2}{k}\delta_{mn}
\end{eqnarray}
where 
\begin{eqnarray}
q_n &=& \alpha_n, n\geq 0, \\
        &=& \bar{\beta}_{-n}, n \leq -1
\end{eqnarray}
This is just the affine $U(1)$ Kac-Moody algebra. 

\section{Conclusions}
In this paper we have studied quantum Hall systems that are coupled through
an interaction at one point. We find that the interaction can lead to
current oscillations between the edge currents of the Hall systems. We have
derived the period of these oscillations in terms of a parameter
characterizing the interaction. It would be interesting to investigate
experimentally the possible values of these parameters.  The fact that
these oscillations arise in situations where there is a violation of the
Kirchoff's law applied to the currents at the interaction junction at any
given point of time and the fact that the currents exhibit a characteristic
periodicity seem to indicate that there may be a natural way to associate
a capacitance to the junction point so that this allows one to understand
the violation of Kirchoff's law in conjunction with the origin of
oscillations.  Thus the kind of boundary conditions that we find may be an
effective description of the mesoscopic quantum capacitance found in
\cite{buttiker}, the capacitance arises here from the microscopic details
of the contact established at the junction.

In this paper we have discussed the modification of abelian current
algebras and the fact that the interactions lead to a twisted current
algebra. It would be interesting to study the corresponding modifications
in the case of non-abelian current algebras as it is known that in the
non-abelian case there are restrictions on the values of the allowed twists
and this leads to constraints on parameters controlling the boundary
interactions.

We have assumed the same filling fraction $k$ for both the discs. We have
to work out the analysis similar to the above for different filling
fractions. More importantly, there is a much more general set of boundary
conditions, with many parameters, that preserve chirality, as shown in the
first section. We have put all except one of the parameters to zero in our
detailed work. We have to try to understand how to exploit the more general
boundary conditions and see if there are interesting solutions that are
also physically relevant.

The connection of our work to string theory and D-branes can be found in
\cite{bachas,acny,fradtset} where the relations between open strings moving
in external fields and boundary conditions of the conformal field theories
are established. The connections to twisted current algebras are also
noticed there.  The fact that the boundary conditions are changed by the
boundary interactions discussed here indicates that there may be common
features of our approach with the one involving the ``twist field''
\cite{diveve}. In the conformal field theory context and in the context of
dissipative quantum mechanics, similar boundary interactions were studied
in \cite{ckmy,cff}. The main difference between those discussions and what
we have in this paper is the fact that we are studying a theory of chiral
bosons.

\section{Acknowledgments}
The work of APB, TRG, VJ and SV was supported in part by Department of
Energy, U. S. A. under contract number DE-FG02-ER40231. VJ would like to
acknowledge The Institute of Mathematical Sciences, Chennai for hospitality
during his visit to Chennai where part of this work was completed.

\bibliographystyle{unsrt}
\bibliography{oscillations}

\end{document}